# The visualization of the space probability distribution for a particle moving in a double ring-shaped Coulomb potential

Yuan You[a)], Fa-Lin Lu [a)], Dong-Sheng Sun [a)], Chang-Yuan Chen [a)], and Shi-Hai Dong[b) 1]

[a)] New Energy and Electronic Engineering, Yancheng Teachers University, Yancheng, 224002, China

[b)] Laboratorio de Información Cuántica, CIDETEC, Instituto Politécnico Nacional, Unidad Profesional Adolfo López Mateos, CDMX, C.P. 07700, Mexico

**Abstract**

The analytical solutions to a double ring-shaped Coulomb potential (RSCP) are presented. The visualizations of the space probability distribution (SPD) are illustrated for the two-(contour) and three-dimensional (isosurface) cases. The quantum numbers ($n$, $l$, $m$) are mainly relevant for those *quasi* quantum numbers ($n'$, $l'$, $m'$) via the double RSCP parameter $c$. The SPDs are of circular ring shape in spherical coordinates. The properties for the relative probability values (RPVs) $P$ are also discussed. For example, when we consider the special case ($n$, $l$, $m$)=(6, 5, 0), the SPD moves towards two poles of axis $z$ when the $P$ increases. Finally, we discuss the different cases for the potential parameter $b$ which is taken as negative and positive values for $c>0$. Compared with the particular case $b=0$, the SPDs are shrunk for $b=-0.5$ while spread out for $b=0.5$.

## 1 Introduction

Since the ring-shaped non-central potentials (RSNCPs) are used to describe the molecular structure of Benzene as well as the interaction between the deformed nucleuses, they have attracted much attention to many authors [1-14]. Generally, these RSNCPs are chosen as the sum of the Coulomb or harmonic oscillator and the single

---

[1] Corresponding authors: Y. You (yuanyou_w@163.com); C.Y. Chen (yctcccy@163.net); S.H. Dong (dongsh2@yahoo.com).

.

ring-shaped part $1/(r^2\sin^2\theta)$ or the double ring-shaped part $1/(r^2\sin^2\theta)+1/(r^2\cos^2\theta)$. In this work, what we are only interested in is the double RSCP, which may be used to describe the properties of ring-shaped organic molecule. The corresponding bound states were investigated by SUSY quantum mechanics and shape invariance [15]. Recently, other complicated double RSCPs have also been proposed [16-25]. Many authors have obtained their solutions in Refs. [7, 8, 13, 14, 26]. Among them the SPDs have been carried out, but their studies are either treated for the radial part in spherical shell $(r, r+dr)$ or for the angular parts [27, 28]. The discussions mentioned above are only concerned with one or two of three variables $(r,\theta,\varphi)$. To show the SPD in all position spaces, we have studied the SPD of a single RSCP for two- and three-dimensional visualizations [29]. In this work, our aim is to focus on the more comprehensive SPD for the particle moving in a double RSCP.

The plan of this paper is as follows. We present the exact solutions to the system in Sec. 2. In Sec. 3 we apply the SPD formula to show the visualizations using the similar technique in [29] getting over the difficulty appearing in the calculation skill when using MATLAB program. We discuss their variations on the number of radial nodes, the RPV $P$ and the RSCP parameter $b$ (positive and negative) when $c \neq 0$ in Sec. 4. We give our concluding remarks in Sec. 5.

## 2. Exact solutions to a double RSCP

In the spherical coordinates the double RSCP is given by

$$V(r,\theta) = -\frac{Ze^2}{r} + \frac{\hbar^2}{2Mr^2}\left(\frac{b}{\sin^2\theta} + \frac{c}{\cos^2\theta}\right) \qquad (1)$$

as plotted in Figs. 1-3.

.

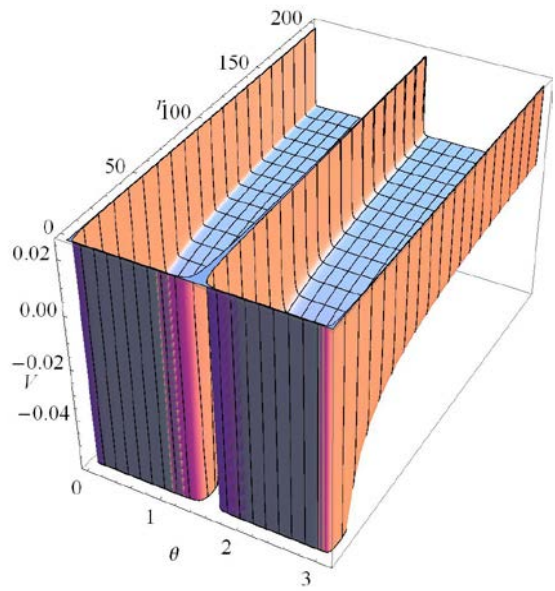

Fig. 1 (Color online) The $V(r,\theta)$ as the functions of variables $\theta$ and $r$.

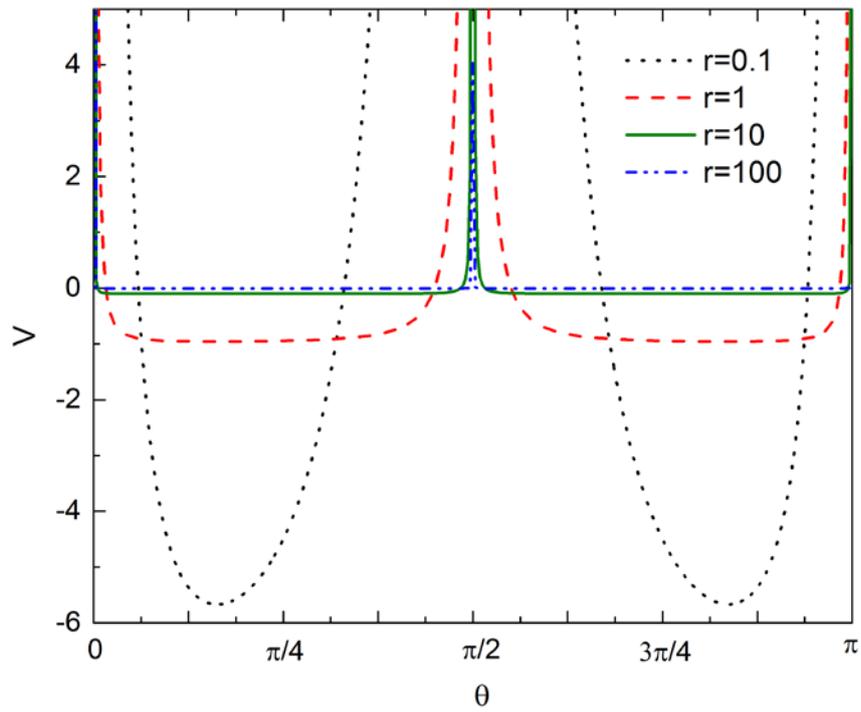

Fig. 2 (Color online) potential function $V(r,\theta)$ versus $\theta$ at $r=0.1, 1, 10, 100$.

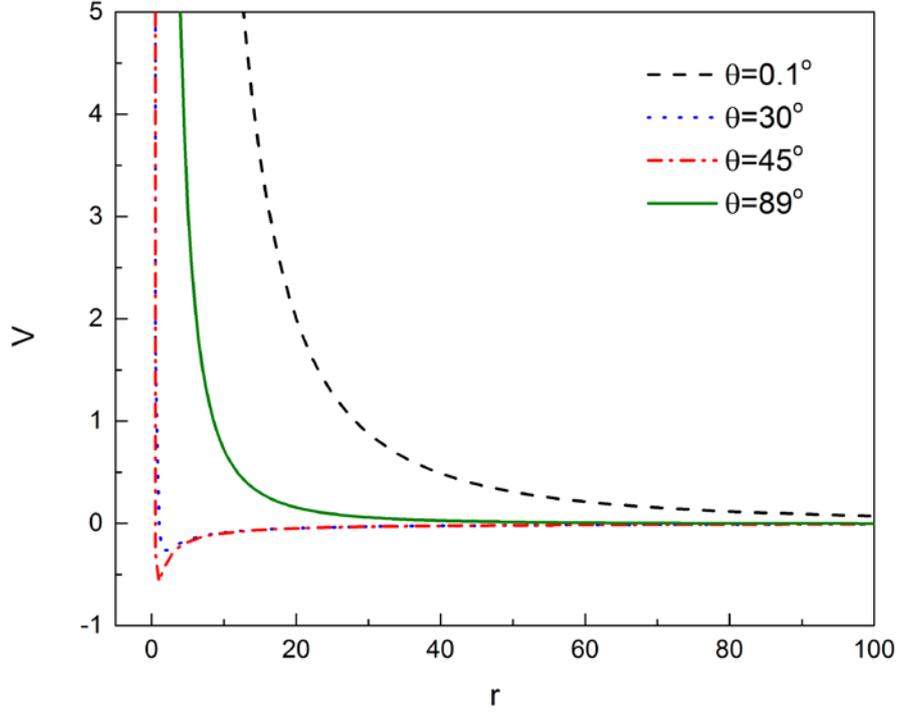

Fig. 3 (Color online) potential function $V(r,\theta)$ versus $r$ at $\theta = 0.1°, 30°, 45°, 89°$.

The Schrödinger equation with this potential is written as ($\hbar=M=e=1$)

$$\left[-\frac{1}{2}\nabla^2 - \frac{Z}{r} + \frac{1}{2r^2}\left(\frac{b}{\sin^2\theta} + \frac{c}{\cos^2\theta}\right)\right]\Psi(\vec{r}) = E\Psi(\vec{r}). \quad (2)$$

Take wavefunction as

$$\Psi(\vec{r}) = \frac{1}{\sqrt{2\pi}} \frac{u(r)}{r} H(\theta) e^{\pm im\varphi}, m = 0, 1, 2, \cdots. \quad (3)$$

Substitute this into Eq. (2) and obtain the following differential equations

$$\frac{d^2 u(r)}{dr^2} + \left(2E + \frac{2Z}{r} - \frac{\lambda}{r^2}\right)u(r) = 0, \quad (4a)$$

$$\frac{1}{\sin\theta}\frac{d}{d\theta}\left(\sin\theta \frac{dH(\theta)}{d\theta}\right) + \left(\lambda - \frac{b+m^2}{\sin^2\theta} - \frac{c}{\cos^2\theta}\right)H(\theta) = 0, \quad (4b)$$

where $\lambda$ is a separation constant. Define $x = \cos\theta$, Eq. (4b) is modified as

$$(1-x^2)\frac{d^2 H(x)}{dx^2} - 2x\frac{dH(x)}{dx} + \left(l'(l'+1) - \frac{(m')^2}{1-x^2} - \frac{c}{x^2}\right)H(x) = 0. \quad (5)$$

Its solutions are given by [30]

.

$$H_{l'm'}(x) = N_{l'm'} P_{l'}^{m'}(\gamma_1, x)$$

$$= N_{l'm'}(1-x^2)^{m'/2} x^{\gamma_1} \sum_{v=0}^{k} \frac{(-1)^v \Gamma(k+\gamma_1-v+1)\Gamma(2l'-2v+1)}{2^{l'} v!(k-v)!\Gamma(2k+2\gamma_1-2v+1)\Gamma(l'-v+1)} x^{2k-2v}. \quad (6)$$

where

$$N_{Lm'} = 2^{\gamma_1} \sqrt{\frac{k!(2l'+1)\Gamma(2k+2\gamma_1+1)\Gamma(l'-k+1)}{2\Gamma(l'-k-\gamma_1+1)\Gamma(k+\gamma_1+1)\Gamma(2l'-2k+1)}} \quad (7)$$

and

$$m' = \sqrt{b+m^2}, l' = n_\theta + m' = 2k + \gamma_1 + \sqrt{b+m^2}, \lambda = l'(l'+1), |m|, k = 0,1,2,\cdots,$$

$$\gamma_1 = \begin{cases} 0 \text{ or } 1, & c = 0 \\ (1+\sqrt{1+4c})/2, & c > 0 \end{cases}. \quad (8)$$

We are now in the position to consider Eq. (4a). Substituting $\lambda = l'(l'+1)$ into (4a) and taking $\chi = \tau r$, $s = 2Z$ and $\tau = Z\sqrt{-1/(2E)}$, then we have from Eq.(4a) as

$$\frac{d^2 u(\chi)}{d\chi^2} + \left(\frac{s}{\chi} - \frac{1}{4} - \frac{l'(l'+1)}{\chi^2}\right) u(\chi) = 0, \quad (9)$$

whose solutions are given by [18]

$$u_{n'l'}(r) = \frac{1}{\Gamma(2l'+2)} \left[\frac{Z}{a_0} \frac{\Gamma(n'+l'+1)}{n_r!(n')^2}\right]^{1/2} \left(\frac{2Zr}{a_0 n'}\right)^{l'+1} e^{-\frac{Zr}{a_0 n'}} F\left(-n_r, 2l'+2, \frac{2Zr}{a_0 n'}\right), \quad (10)$$

where

$$n' = n_r + l' + 1 = \begin{cases} n_r + 2k + m' + (3+\sqrt{1+4c})/2, & c > 0, n_r, m, k = 0,1,2,\cdots \\ n_r + k + m' + 1, & c = 0, n_r, m, k = 0,1,2,\cdots \end{cases},$$

and the Bohr radius $a_0 = \hbar^2/Me^2 = 1$. The complete wavefunction has the form

$$\Psi_{n'l'm}(\vec{r}) = \frac{1}{\sqrt{2\pi}} \frac{u_{n'l'}(r)}{r} H_{l'm'}(\cos\theta) e^{\pm im\varphi}. \quad (11)$$

## 3. Two- and three dimensional visualizations of SPDs

As we know, the SPDs at the position $\vec{r} = (r, \theta, \varphi)$ are calculated by

$$\rho = |\Psi_{n'l'm'}(\vec{r})|^2 = \frac{1}{2\pi} \frac{u_{n'l'}^2(r)}{r^2} H_{l'm'}^2(\cos\theta). \quad (12)$$

.

To show the SPD let us transform Eq. (12) to popular Cartesian coordinates via the relations $r = \sqrt{x^2 + y^2 + z^2}$, $\cos\theta = z/r$. Thus, one is able to find the corresponding SPD $\rho(x, y, z)$.

Taking a series of discrete positions we may study the values of the respective SPD by numerical calculation. In order to make the graphic resolution better, one takes $N$ discrete positions in the Cartesian space ($x, y, z$) and studies density block, say den($N, N, N$), which is composed of all values $w_{n'l'm'}$ for all $N \times N \times N$ positions. Here, the $N$ is taken as 151. For state denoted by $(n', l', m')$, we display their two-and three-dimensional visualizations for different states ($n \leq 6$) using MATLAB program as shown in Tables 1 and 2.

## 4. Discussions on the SPD

### 4.1 *Variation caused by the radial nodes*

We show the SPDs for various cases $b = 0.5$, $c = 0, 0.5, 5$ (see In Table 1). The case $c=0$ corresponds to a single RSCP, which was discussed in our previous works [29, 30]. We take the unit in axis as the Bohr radial $a_0$. It should be pointed that we plot the Figures only for the value of $n_r = n - l - 1$ equal to integer. To display the inside structure of the graphics, we create a section plane but need not consider their numerical values in the regions $x < 0, y < 0, z > 0$.

Compared to the cases $c = 0$ and $b \neq 0$, it is seen that the graphics are expanded. That is to say, the SPDs enlarge toward axis $z$ and the hole is expanded outside when the potential parameter $c$ increases. We may understand it through considering Eq. (8). As we know, the $m'$ will increase relatively for a fixed $m$. For $l \neq m$, their isosurfaces are of circular ring shape.

We project the SPD to a plane $yoz$ and find that they are symmetric to the axis $y$ and $z$ (see Table 2). In this work, the graphics are plotted only in the first

.

quadrant by enlarging proportionally the probability $|\Psi_{n'l'm'}(\vec{r})|^2$ and by making the maximum value as 100, in which the interval is taken as 10. A corresponding balance among the density distributions exists in the directions of axis *x*, *y* and *z* because the sum of density distributions has to be equal to one when considering the normalization condition. It is clearly that each figure becomes expanded along with the *y*- axis and *z*- axis.

### 4.2 *Variation on RPV P*

To show the isosurface of the SPDs for various RPV $P \in (0,100)\%$, the quantum numbers (*n*, *l*, *m*)=(6, 5, 0) for two different cases $b = 0.5$, $c = 0.5, 10$ are taken as see in Table 3. It is shown that the particle for the smaller *P* will be distributed to almost all spaces. However, the particle for the larger *P* will move to the poles in the *z*- axis.

### 4.3 *Variations on various potential parameters b and c*

Considering given quantum numbers $m$ and $n_\theta$, we know from Eq. (8) that the $m'$ will become bigger as the *b increases* and the parameter $\gamma_1$ also becomes larger with increasing *c*. As a result, this will result in the increasing $l'$. In Table 4 the SPDs are plotted for state $(n,l,m) = (5,1,0)$ in the cases of $c = 0.5, b = 0, 5, 10, 25, 40, 80$ and $b = 0.5, c = 0, 5, 10, 25, 40, 80$, respectively. Obviously, we see a big difference between them. When the potential parameter *b* increases, the expansions of the SPDs along with the *x*- axis and *y*-axis and the number of radial nodes are changed. However, when the parameter *c* increases, the expansions of the SPDs are along with the *z*- axis.

  The comparison is done for positive and negative $b < 0$ and $b > 0$ when $c = 0.5$ (see Table 5). It is shown that the SPDs for the negative $b = -0.5$ compared with the case $b=0$ are shrunk into the origin. However, the SPDs for $b=0.5$ are enlarged outside. We can understand it very well by studying the contributions of the

.

potential parameter *b* made on the Coulomb potential. The choice of the negative or positive *b* determines the attractive Coulomb potential which is bigger or smaller relatively. Thus, the attractive force which acts on the particle will be larger or smaller. Finally, this will result in the SPDs which are shrunk or expanded.

## 5. Conclusions

The analytical solutions to the double RSCP have been obtained and then the visualization of the SPDs for this potential are performed. The contour and isosurface visualizations have been illustrated for quantum numbers $(n',l',m')$ by taking various values of the parameter *c*. It is shown that the SPDs are of circular ring shape. On the other hand, the properties of the RPVs *P* of the SPDs. As an example, we have studied the particular case, that is, (*n, l, m*)=(6, 5, 0) and found that the SPDs will move towards to the poles of axis *z* when the RPVs *P* increase.

**Acknowledgments**: This work is supported by the NNSF of China under Grant No. 11275165, partially by 20180677-SIP-IPN and the CONACYT under Grant No. 288856-CB-2016, Mexico. Prof. You thanks Jiangsu Overseas research & Training Program for University Prominent Young & Middle-aged Teachers and Presidents to support.

.

.

.

Table 1: The isosurface SPDs with a section plane

| *n* | *l* | *M* | b=0.5<br>c=0 | b=0.5<br>c=0.5 | b=0.5<br>c=5 |
|---|---|---|---|---|---|
| 2 | 1 | 0 | 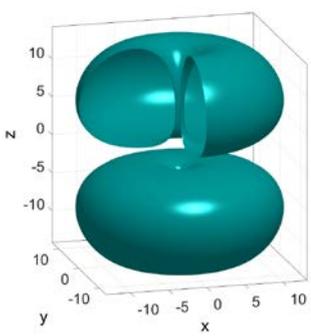 | 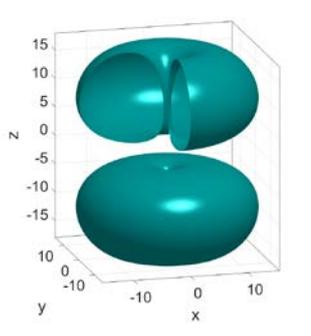 | 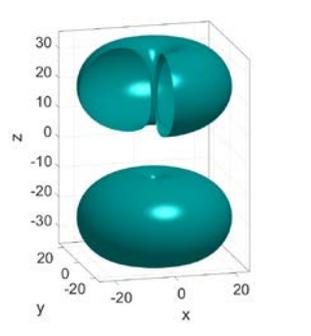 |
| 3 | 1 | 0 | 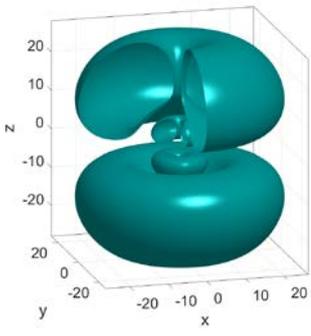 | 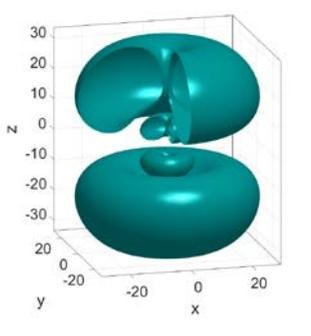 | 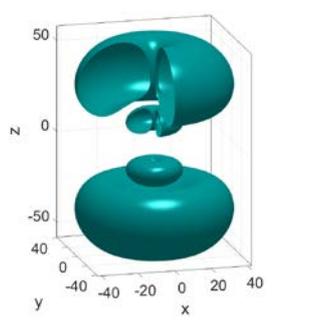 |

| 3 | 2 | 1 | 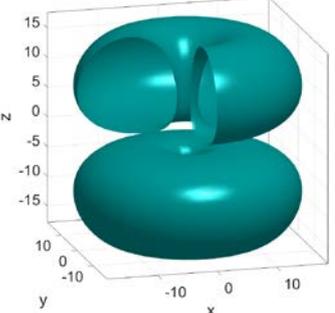 | 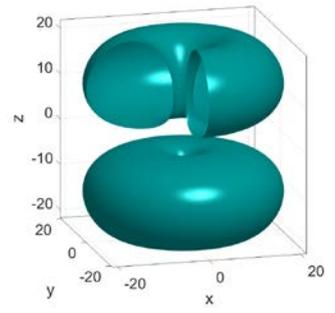 | 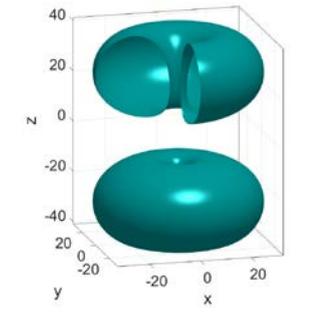 |
|---|---|---|---|---|---|
| 4 | 1 | 0 | 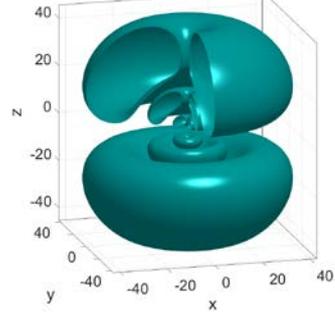 | 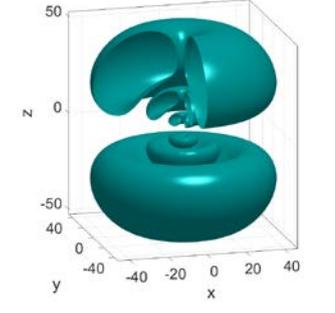 | 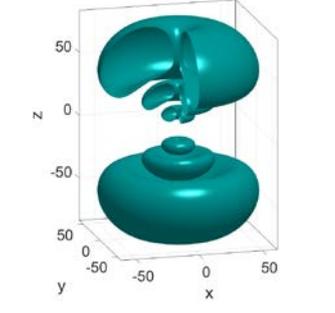 |
| 4 | 2 | 1 | 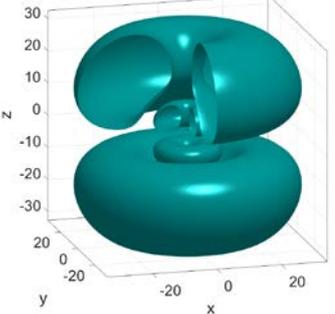 | 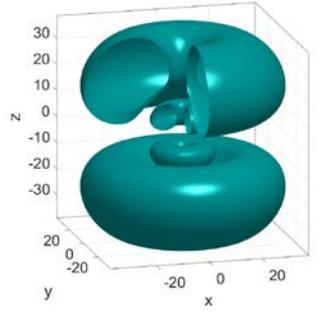 | 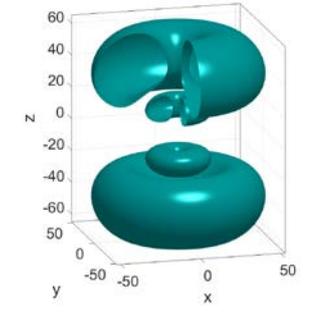 |
| 4 | 3 | 0 | 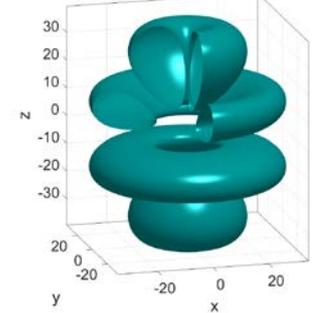 | 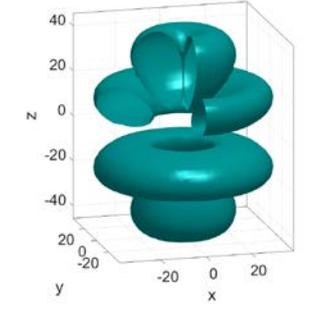 | 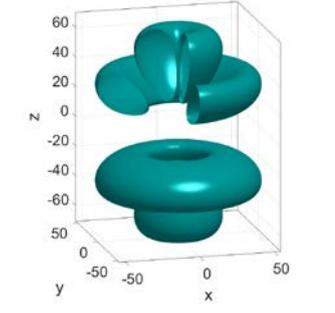 |

| 4 | 3 | 2 | 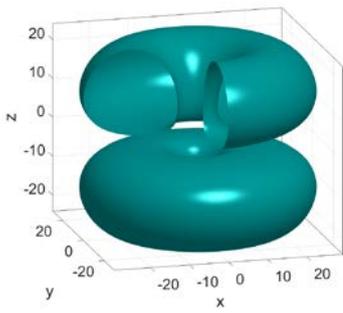 | 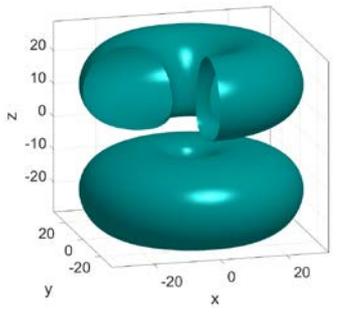 | 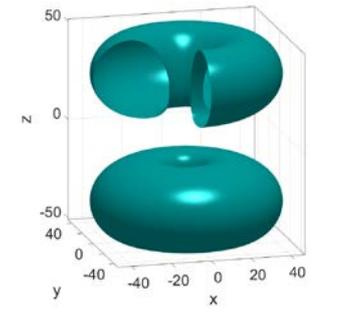 |
|---|---|---|---|---|---|
| 5 | 1 | 0 | 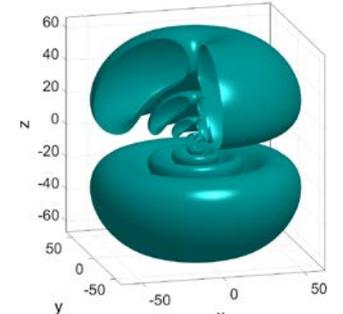 | 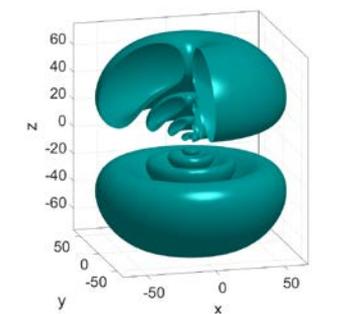 | 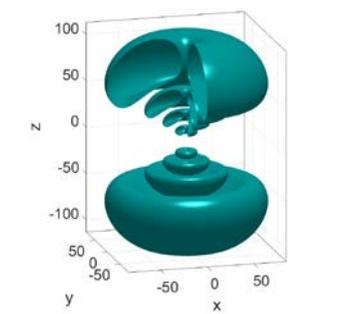 |
| 5 | 2 | 1 | 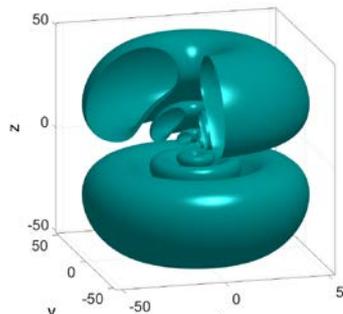 | 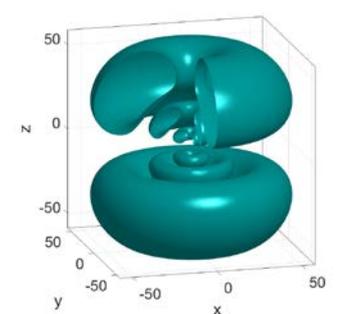 | 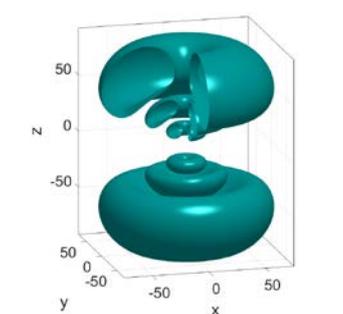 |
| 5 | 3 | 0 | 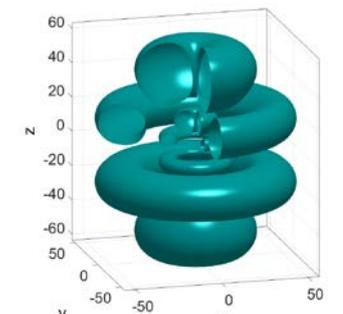 | 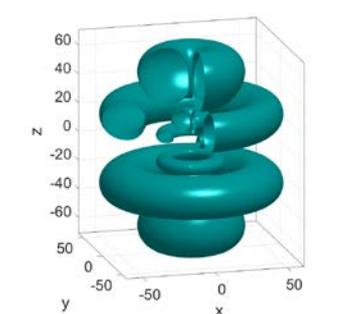 | 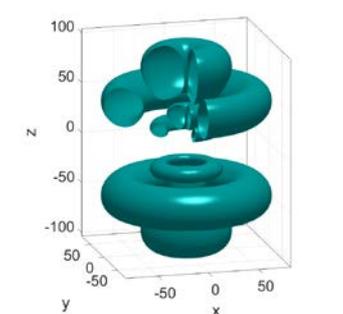 |
| 5 | 3 | 2 | 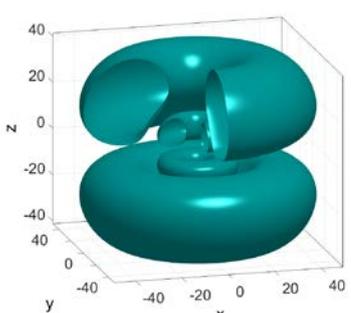 | 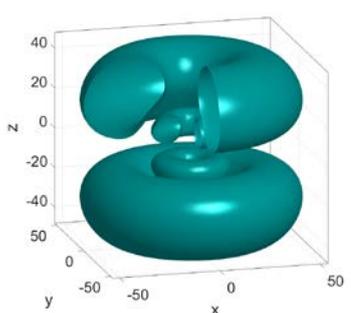 | 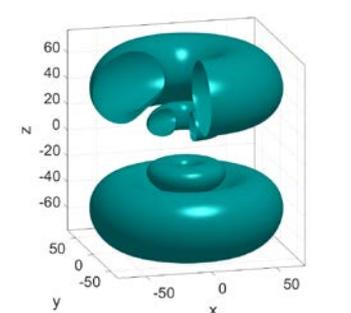 |
| 5 | 4 | 1 | 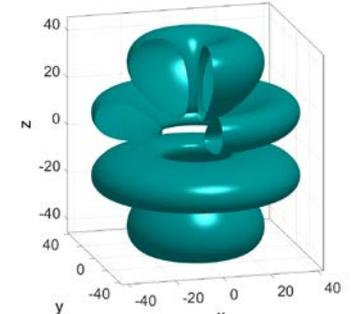 | 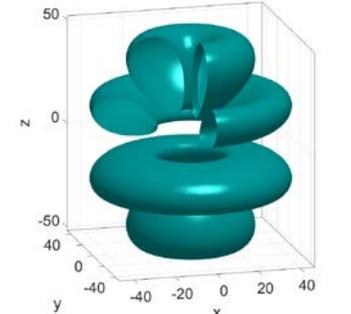 | 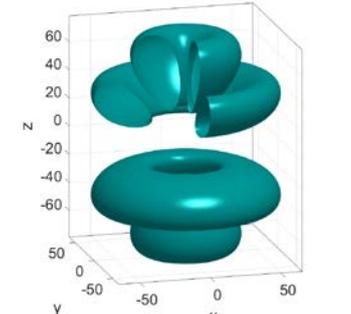 |
| 5 | 4 | 3 | 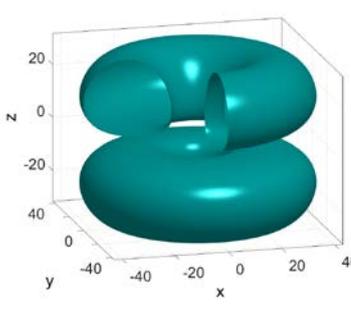 | 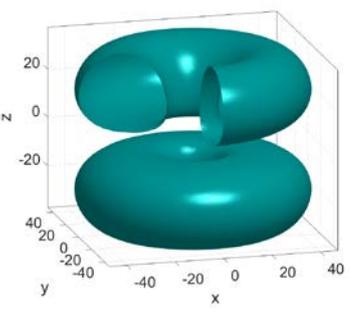 | 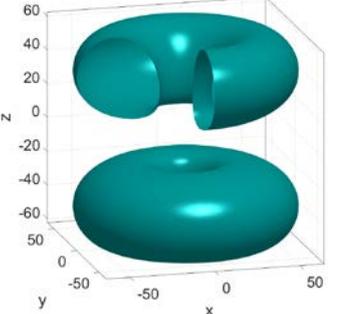 |
| 6 | 1 | 0 | 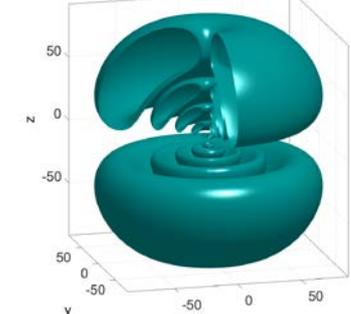 | 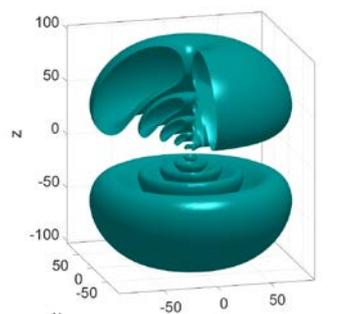 | 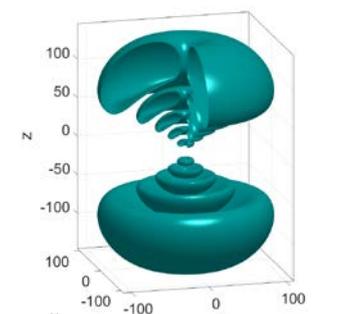 |

| 6 | 2 | 1 | 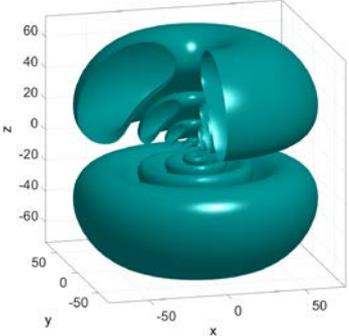 | 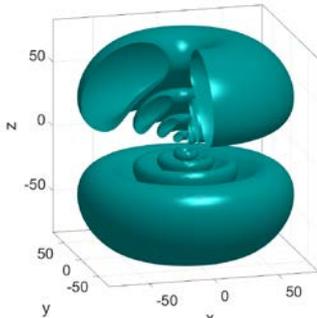 | 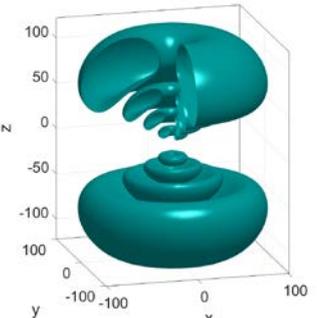 |
|---|---|---|---|---|---|
| 6 | 3 | 0 | 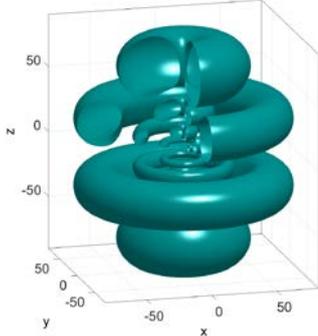 | 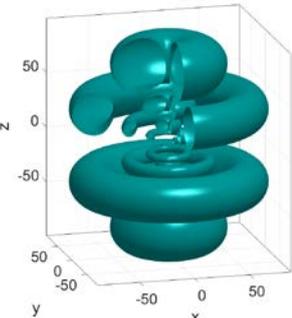 | 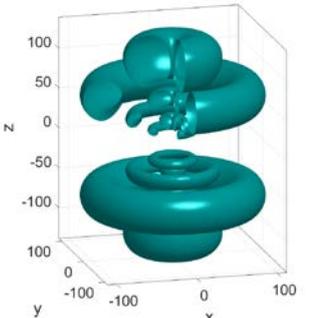 |
| 6 | 3 | 2 | 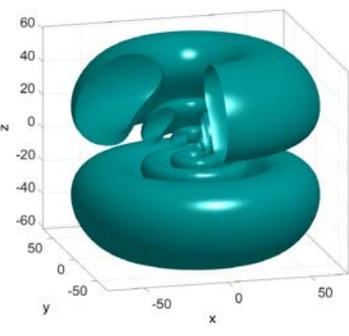 | 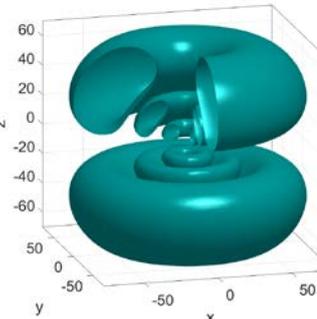 | 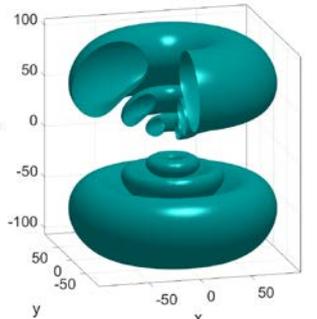 |
| 6 | 4 | 1 | 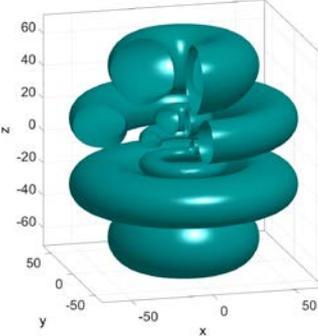 | 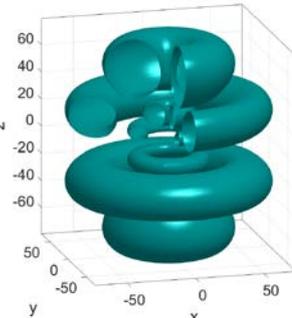 | 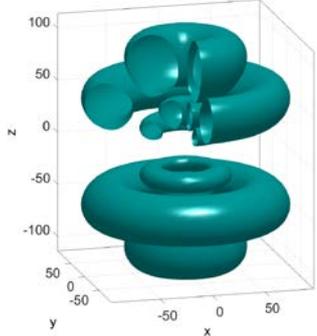 |

| 6 | 4 | 3 | 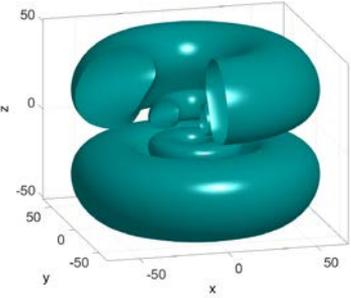 | 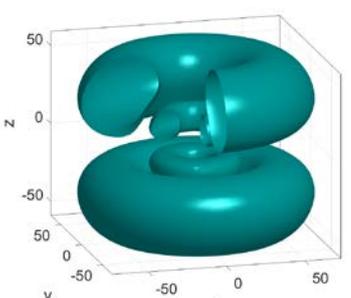 | 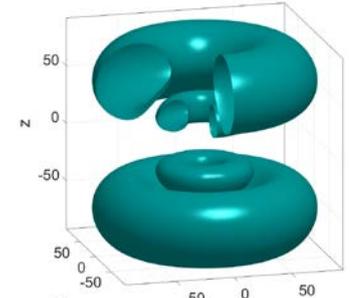 |
|---|---|---|---|---|---|
| 6 | 5 | 0 | 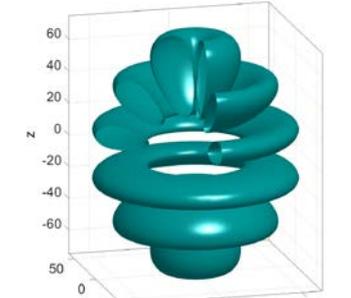 | 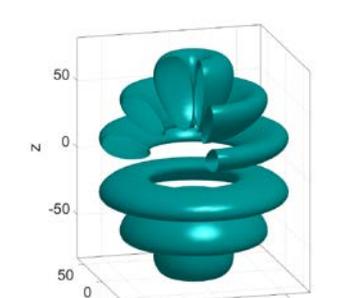 | 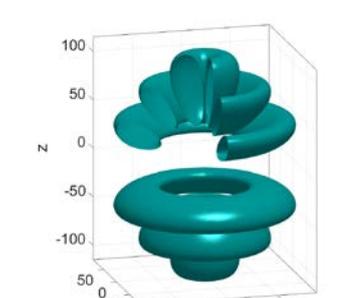 |
| 6 | 5 | 2 | 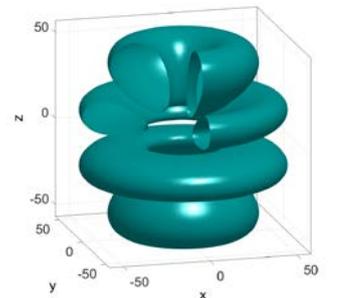 | 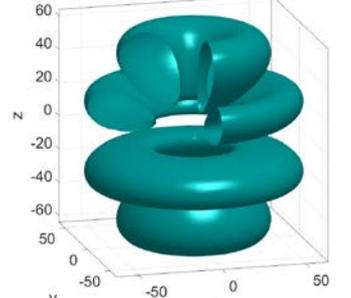 | 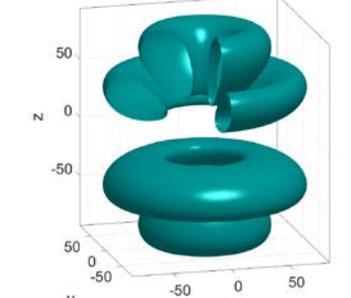 |
| 6 | 5 | 4 | 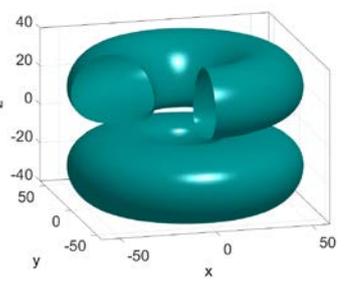 | 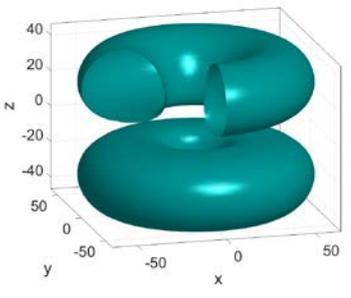 | 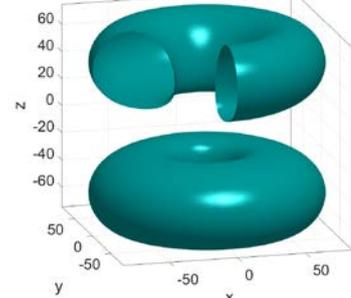 |

Table 2: The Contour of the SPDs in the plane *yoz*

| $n$ | $l$ | $m$ | **b=0.5** **c=0** | **b=0.5** **c=0.5** | **b=0.5** **c=5** |
|---|---|---|---|---|---|
| 2 | 1 | 0 | | | |
| 3 | 1 | 0 | | | |
| 3 | 2 | 1 | | | |
| 4 | 1 | 0 | | | |

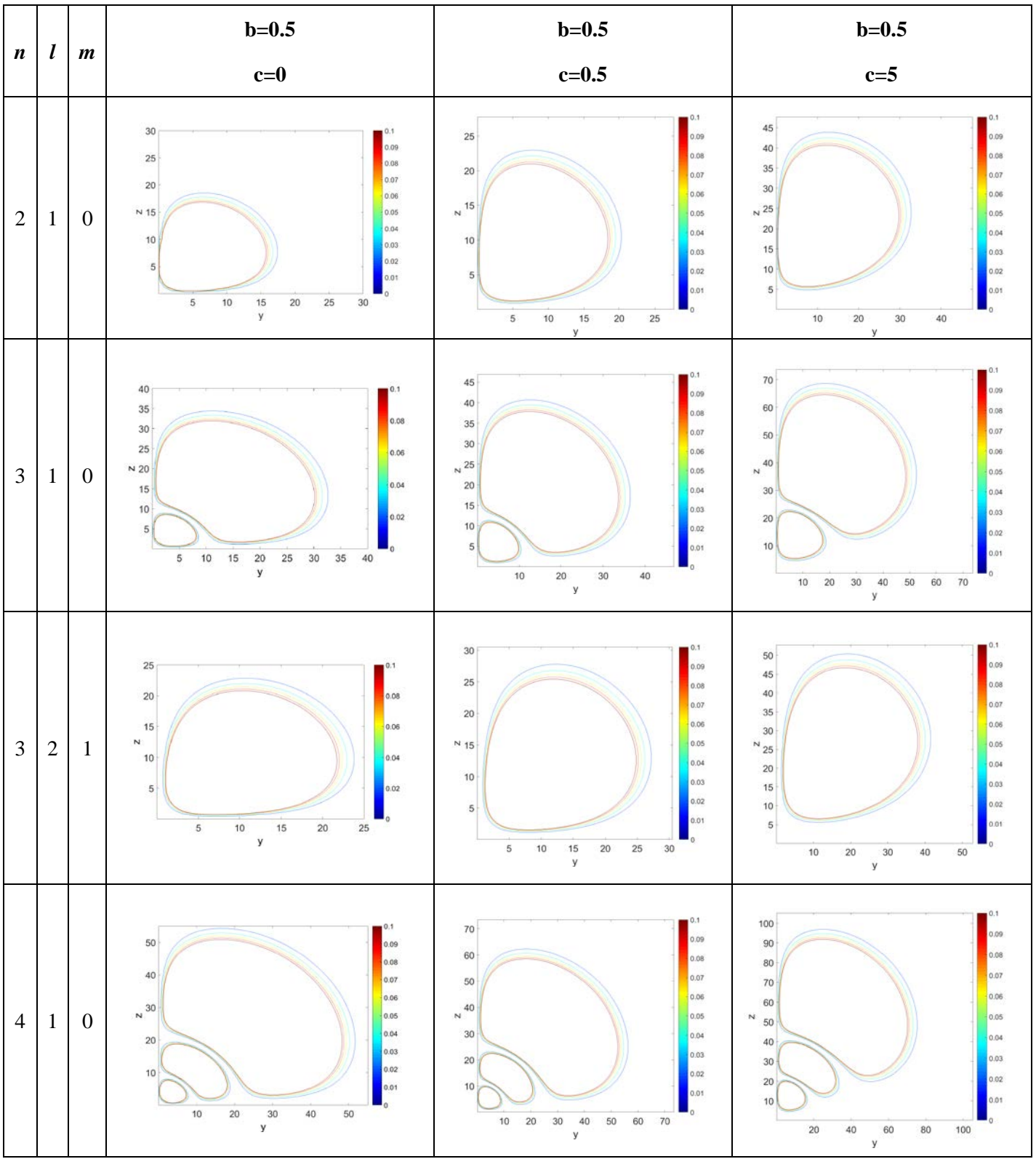

| 4 | 2 | 1 | | | |
|---|---|---|---|---|---|
| 4 | 3 | 0 | | | |
| 4 | 3 | 2 | | | |
| 5 | 1 | 0 | | | |
| 5 | 2 | 1 | | | |

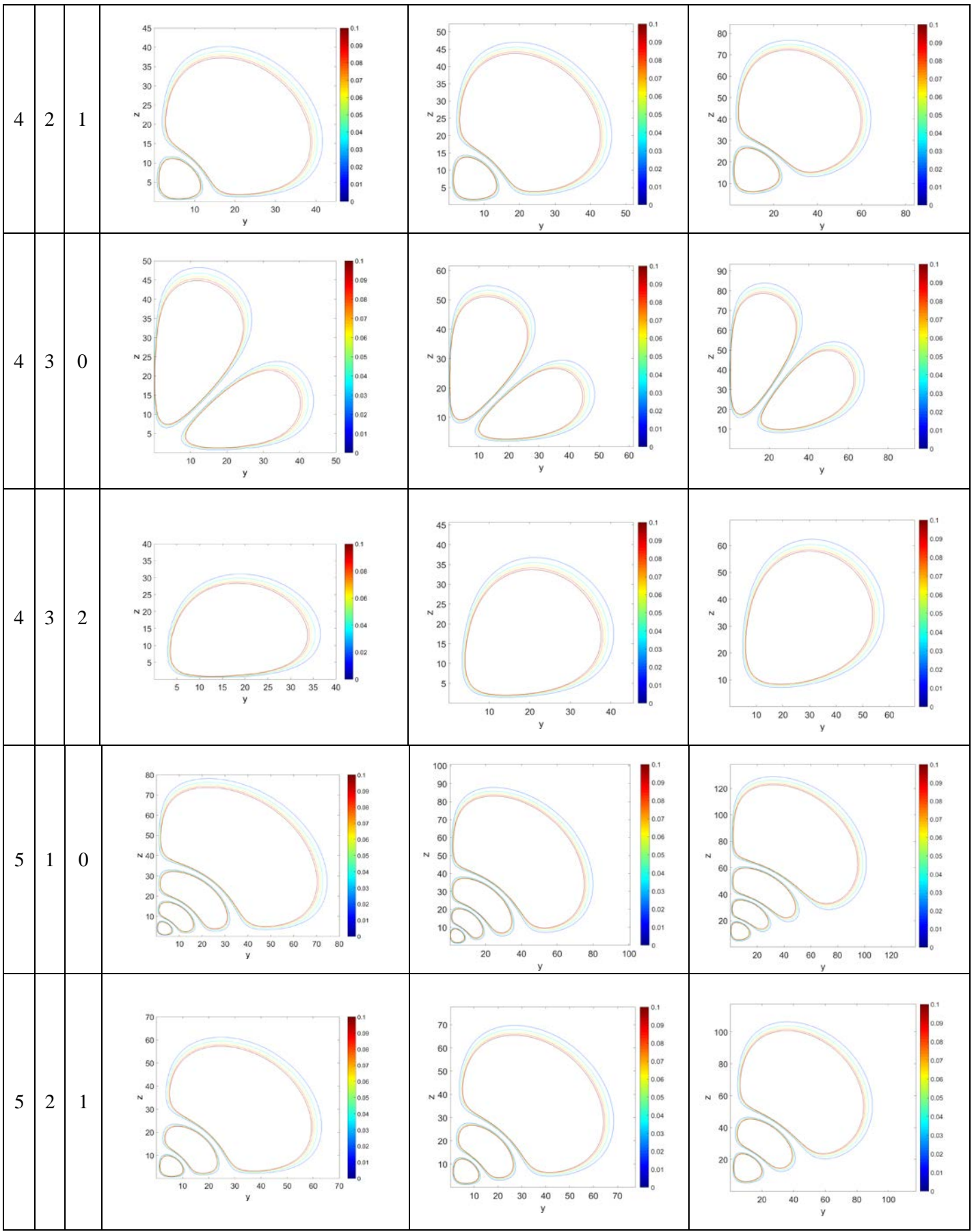

| 5 | 3 | 0 | | | |
|---|---|---|---|---|---|
| 5 | 3 | 2 | | | |
| 5 | 4 | 1 | | | |
| 5 | 4 | 3 | | | |
| 6 | 1 | 0 | | | |

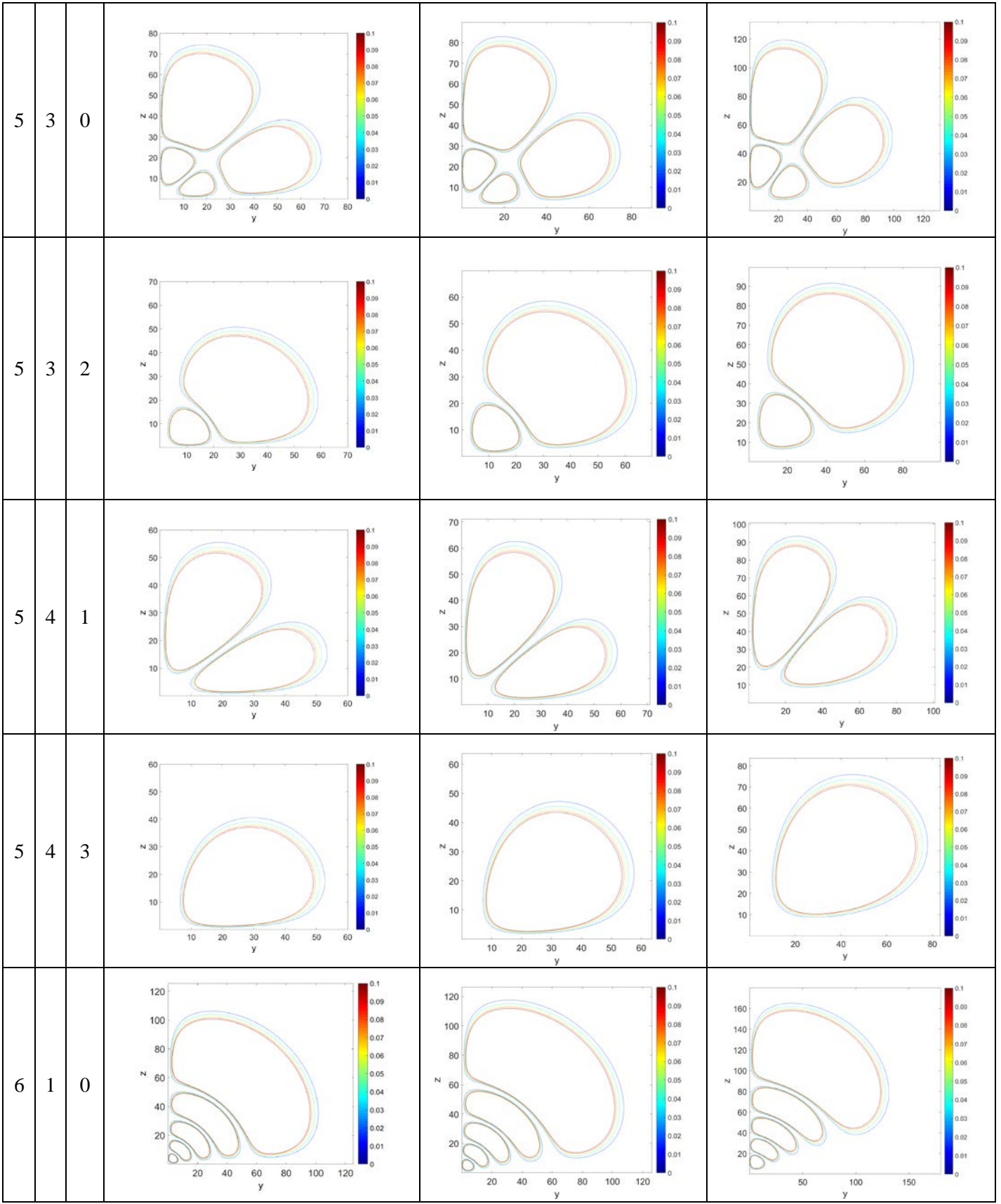

| 6 | 2 | 1 | | | |
|---|---|---|---|---|---|
| 6 | 3 | 0 | | | |
| 6 | 3 | 2 | | | |
| 6 | 4 | 1 | | | |
| 6 | 4 | 3 | | | |

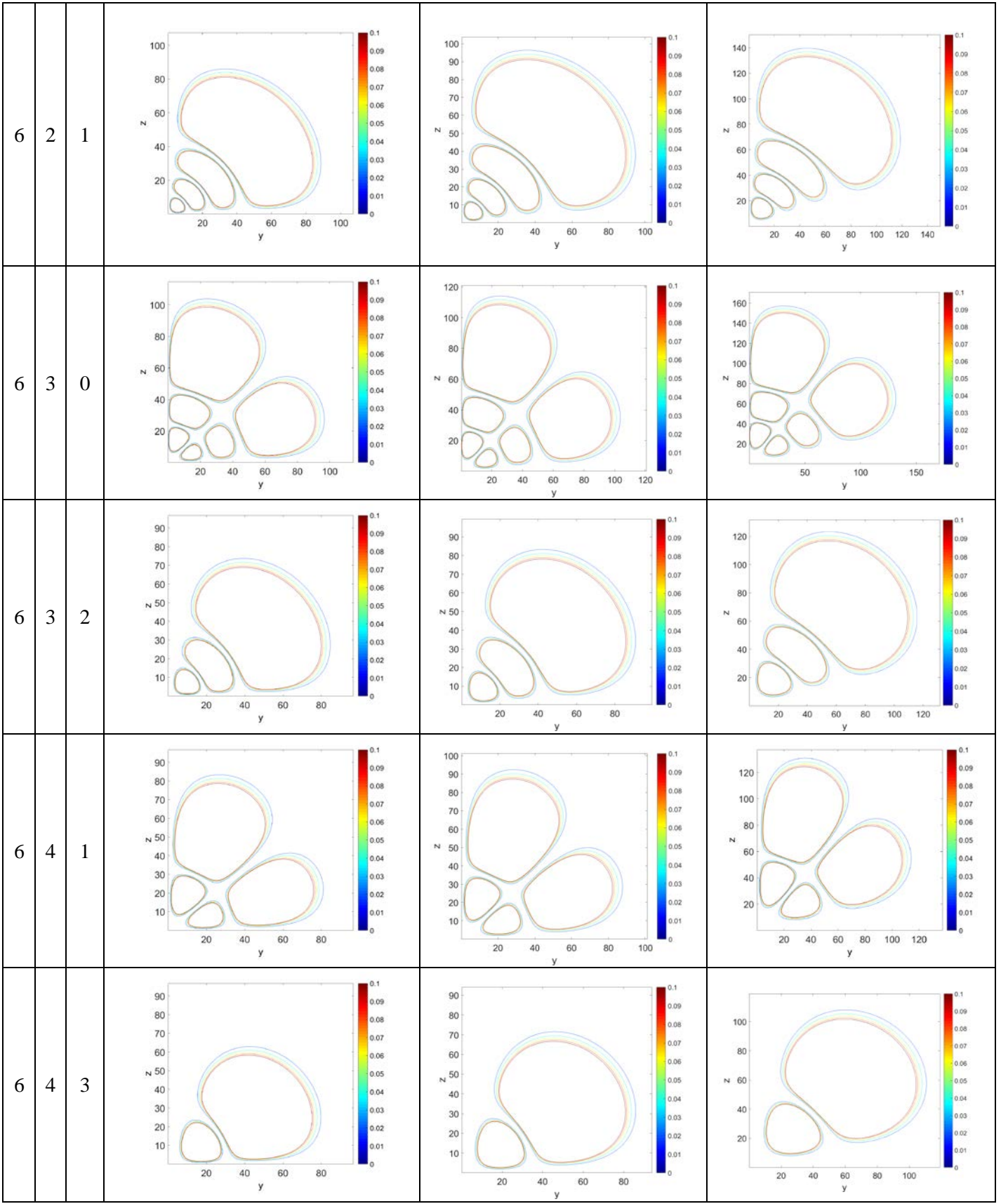

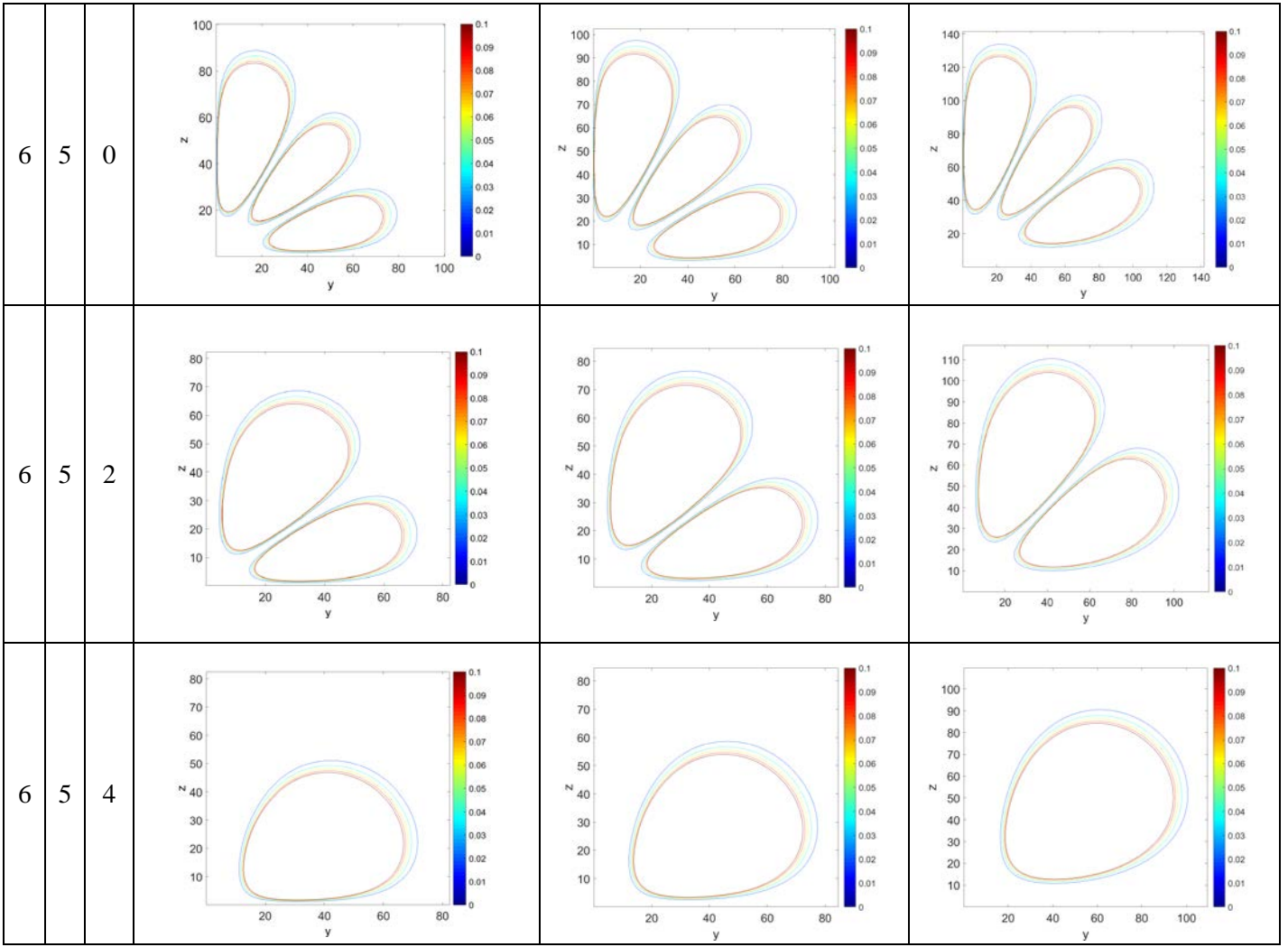

Table 3: The SPDs for various RPVs *P* for (*n*, *l*, *m*)=(6, 5, 0) (b=0.5, c=0.5,10)

| *P* (%) | c=0.5 | c=10 |
|---|---|---|
| 0.01 | 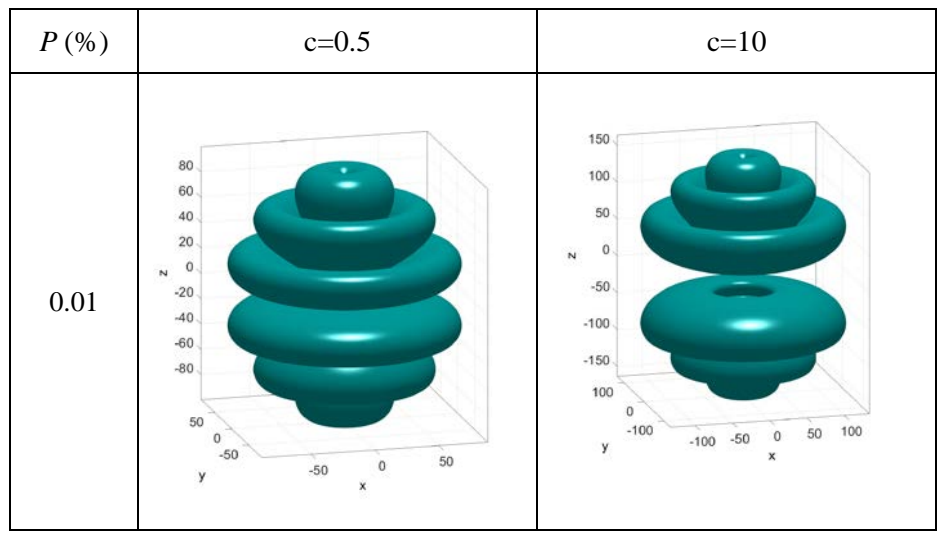 | |

| 1 | | |
|---|---|---|
| 5 | | |
| 10 | | |
| 20 | | |

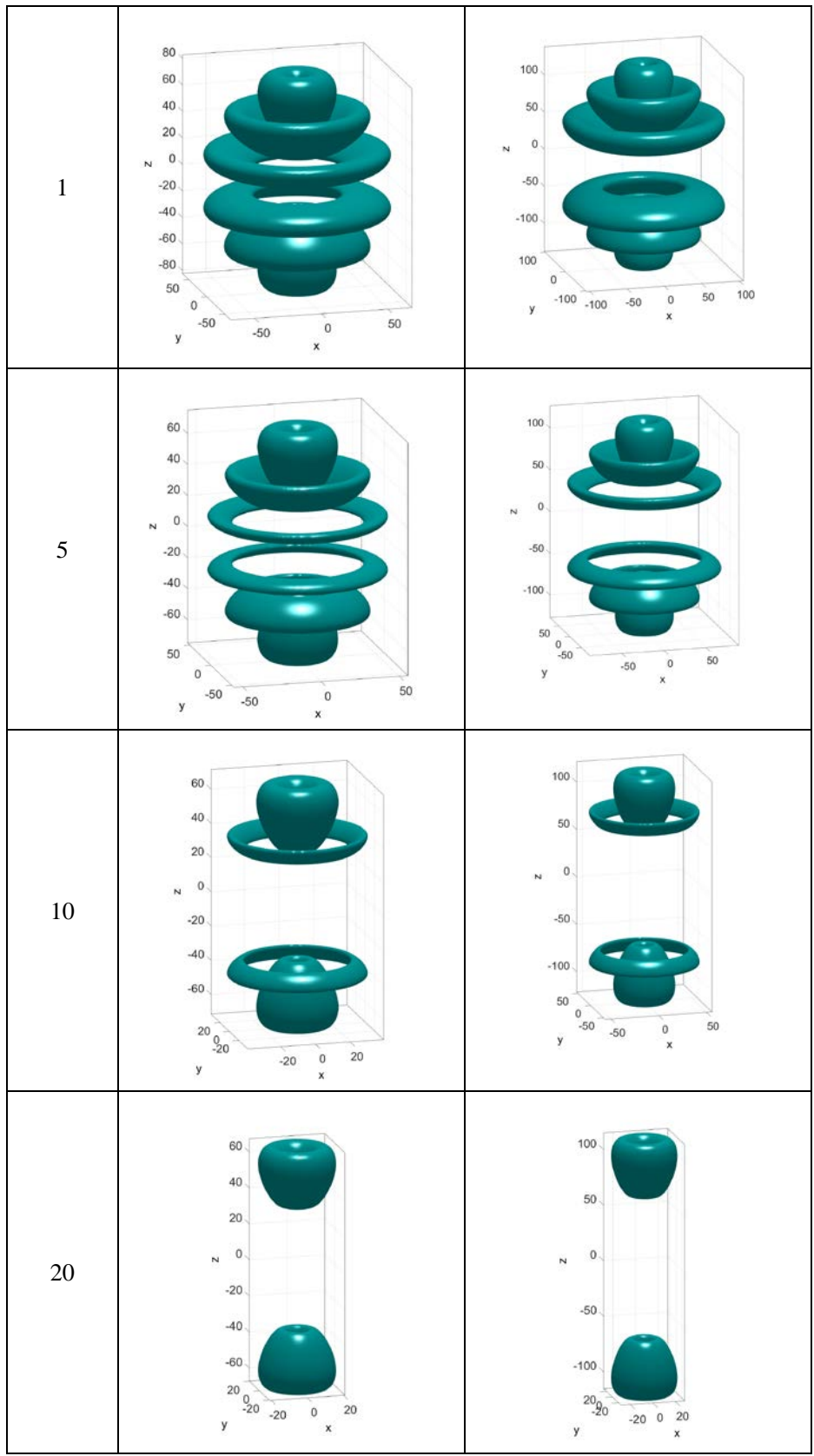

| 50 | 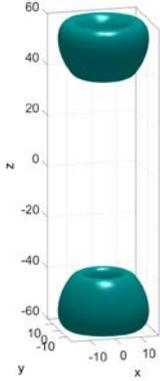 | 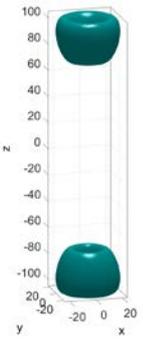 |
|---|---|---|
| 70 | 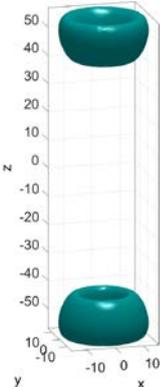 | 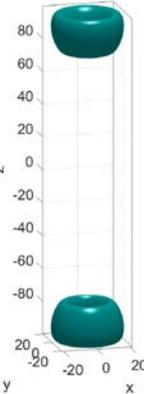 |
| 90 | 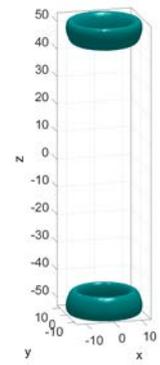 | 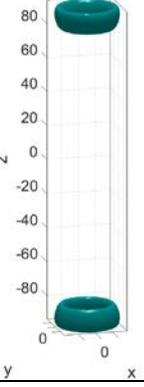 |
| 99 | 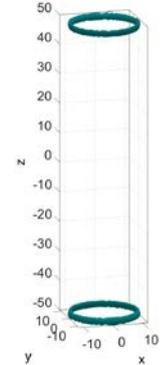 | 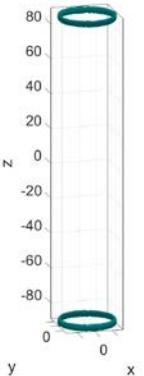 |

Table 4: The isosurface illustration of the state (5, 1, 0) with various values *b and c*

| $b$ (*c*=0.5) | Isosurface illustration | $c$ (*b*=0.5) | Isosurface illustration |
|---|---|---|---|
| **0** | 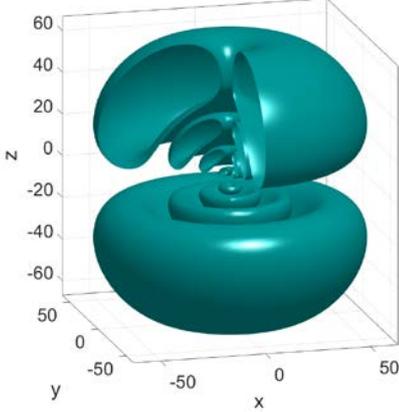 | **0** | 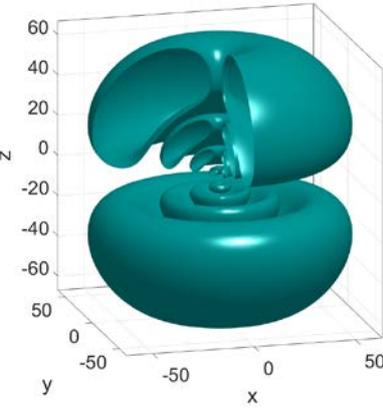 |
| **5** | 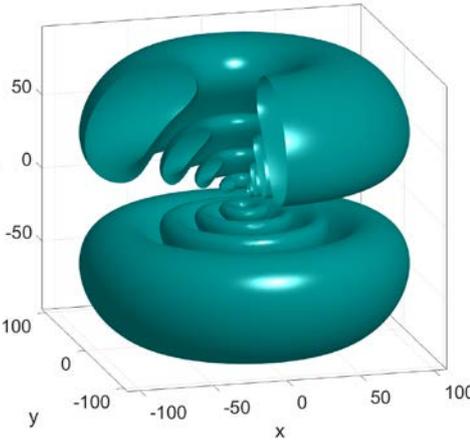 | **5** | 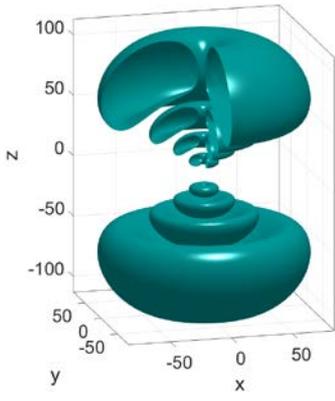 |
| **10** | 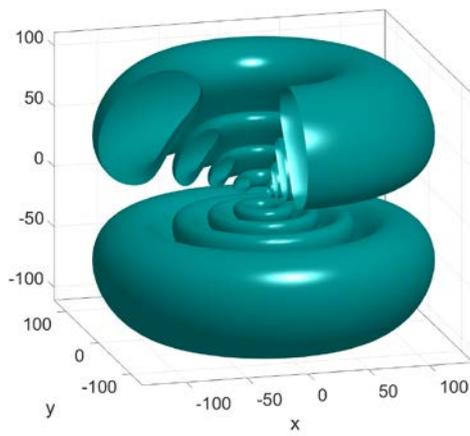 | **10** | 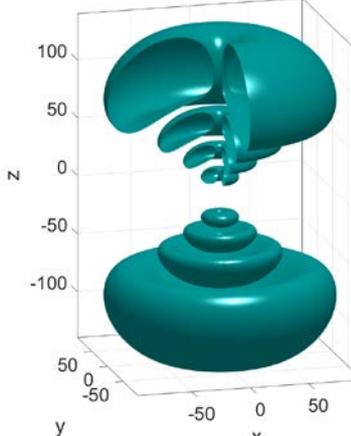 |

| b | Isosurface illustration | | Contour illustration |
|---|---|---|---|
| **25** | 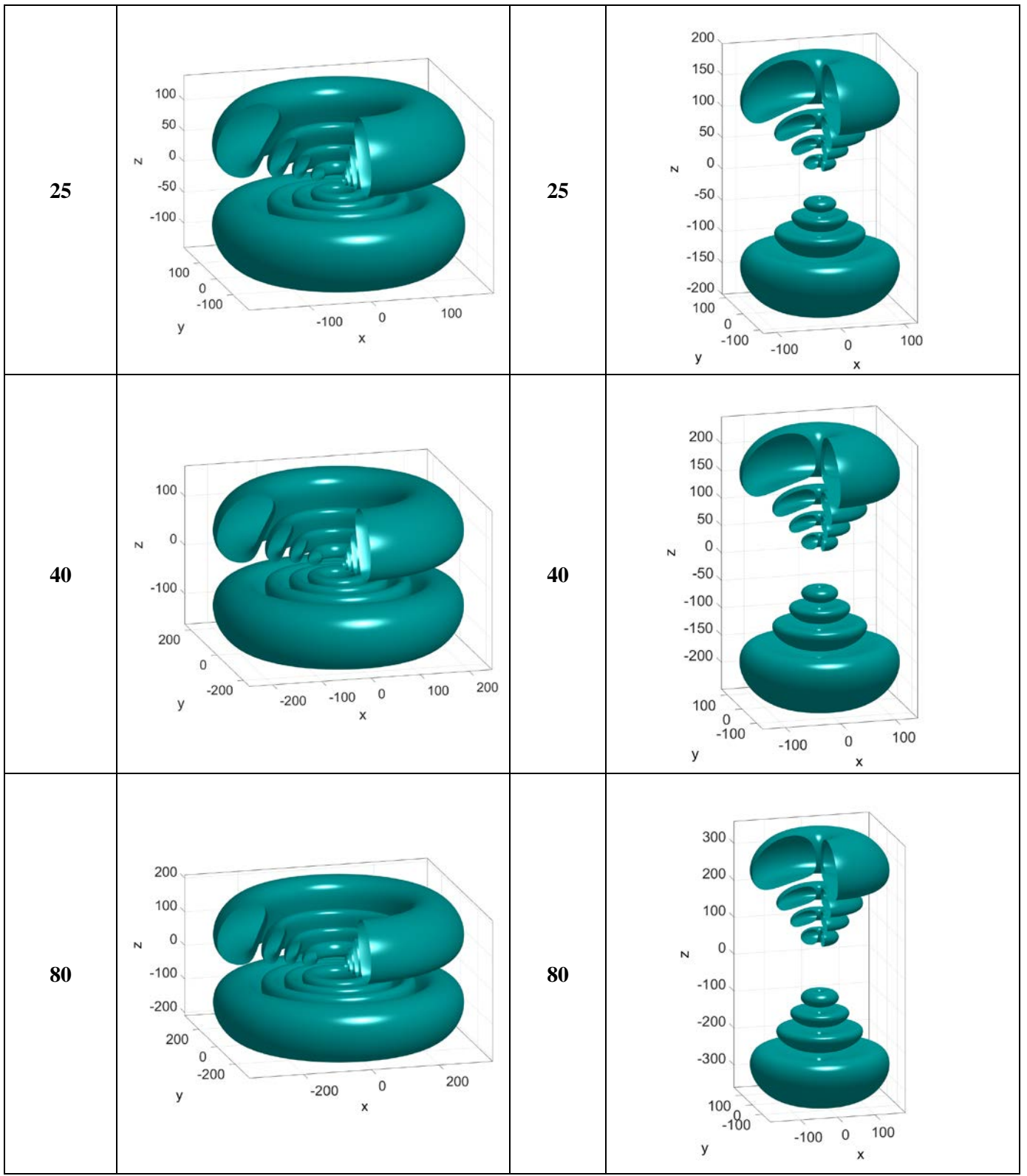 | **25** | |
| **40** | | **40** | |
| **80** | | **80** | |

Table 5: SPDs for different cases of the value *b* for (4, 1, 0) when *c*=0.5

| *b* | Isosurface illustration | Contour illustration |
|---|---|---|

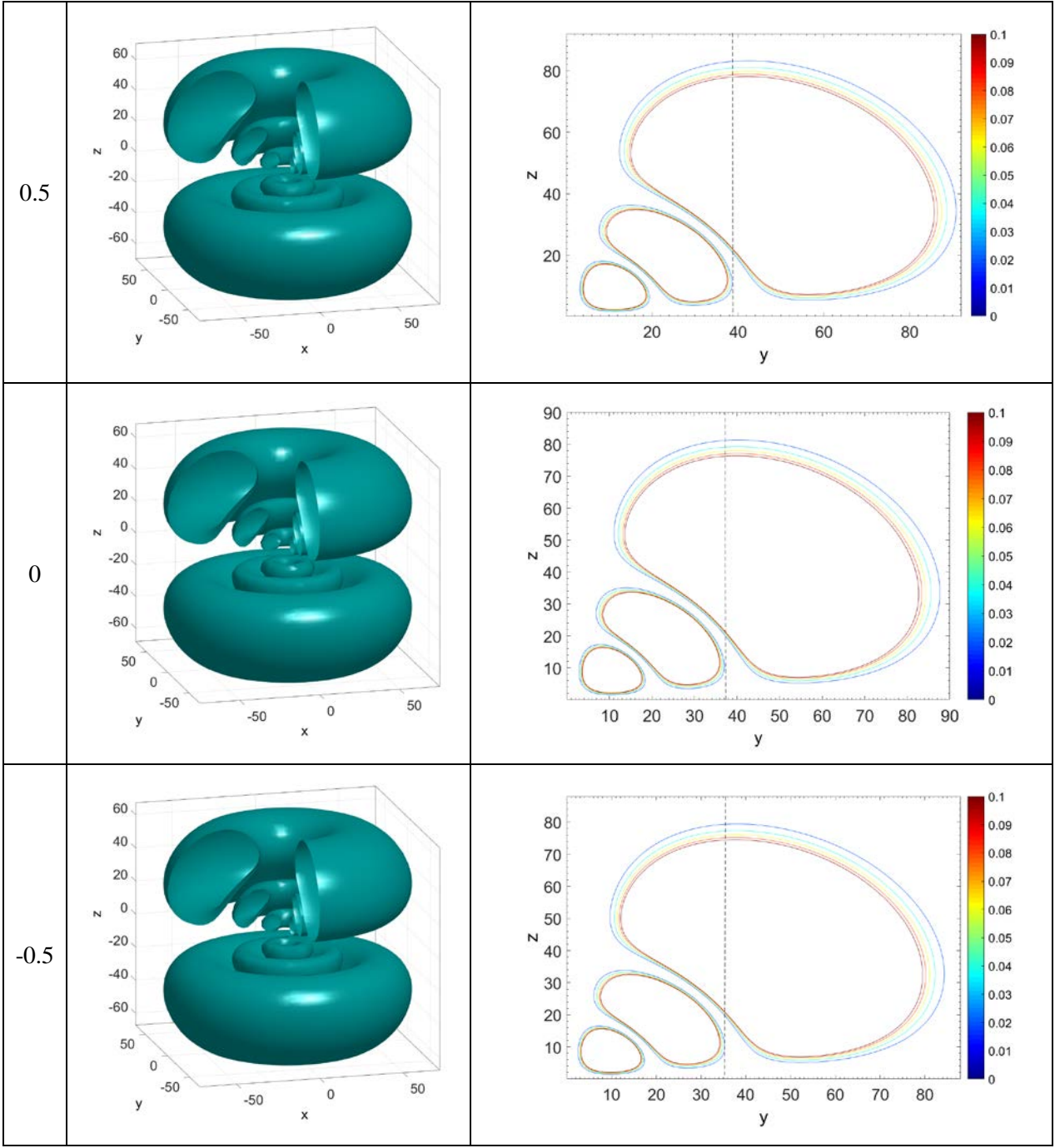